\documentclass[aps,prl,preprint]{revtex4-1} 
\pdfoutput=1
\usepackage{graphicx}
\usepackage{amsmath}
\usepackage{epstopdf}
\usepackage{hyperref}
\usepackage{color}
\begin{document}
\newcommand{\<}{\langle}
\renewcommand{\>}{\rangle}
\newcommand{\beq}{\begin{equation}}
\newcommand{\eeq}{\end{equation}}
\newcommand{\kc}{K_{\text{c}}}
\newcommand{\lt}{L_\tau}
\newcommand{\asp}{\alpha_{\tau}}
\newcommand{\br}{{\bf r}}
\newcommand{\bk}{{\bf k}}
\newcommand{\bJ}{{\bf J}}
\newcommand{\bP}{{\bf P}}
\newcommand{\bh}{{\bf h}}
\newcommand{\bA}{{\bf A}}
\newcommand{\bB}{{\bf B}}

\title{Scaling of the magnetic permeability at the Berezinskii-Kosterlitz-Thouless transition
from Coulomb gas simulations}

\author{Rogelio D\'iaz-M\'endez} 
\affiliation{Department of Physics, KTH Royal Institute of Technology,
  SE-106 91 Stockholm, Sweden}

\author{Jack Lidmar} 
\affiliation{Department of Physics, KTH Royal Institute of Technology,
  SE-106 91 Stockholm, Sweden}

\author{Mats Wallin} 
\affiliation{Department of Physics, KTH Royal Institute of Technology,
  SE-106 91 Stockholm, Sweden}

\date{\today}

\begin{abstract}
  A new approach to the Berezinskii-Kosterlitz-Thouless transition in
  the two-dimensional Coulomb gas model is explored by Monte Carlo simulation
  and finite size scaling.
  The usual mapping of a neutral two-dimensional superconductor in zero 
  magnetic field to a Coulomb gas leads to an
  unscreened logarithmic interaction between the vortices,
  and with periodic boundary conditions
  vortex configurations are always vorticity neutral with an equal
  number of plus and minus vortices. 
  We demonstrate that relaxing the
  neutrality condition has certain advantages. It leads to non-neutral
  vortex configurations that can appear in real systems with open
  boundary conditions and permits calculation of the compressibility,
  which for thin film superconductors corresponds to the magnetic
  permeability.  The vortex-number fluctuation has remarkable scaling
  properties at and below the Berezinskii-Kosterlitz-Thouless transition. The
  fugacity variable becomes dangerously irrelevant in the
  low-temperature phase and leads to a multiplicative scaling
  correction to the mean-square vortex-number fluctuation and to the
  magnetic permeability.
  This multiplicative
  correction strongly affects the scaling properties of the vorticity
  fluctuation at and below the transition.  Consequences of these findings are
  demonstrated using Monte Carlo simulations.  Inclusion of the
  next-higher order correction to scaling is found to play an
  important role in the analysis of numerical data for the vortex
  number fluctuation and permits accurate determination of the
  critical properties.
\end{abstract}

\maketitle

\section{Introduction}

The Berezinskii-Kosterlitz-Thouless (BKT) transition
\cite{Berezinskii1,Berezinskii2,KT} is a paradigm shifting phenomenon
that demonstrates how topological fluctuations can create phase
transitions without the Landau symmetry breaking mechanism. The
topological excitations are quantized vortices and the BKT transition
is a pair-unbinding transition, where the vortex correlations change
long-distance behavior from algebraic to exponentially decaying.  The
BKT transition is found in several experimental ultrathin-film
systems, e.g., superfluids, superconductors, Josephson-junction
arrays, planar magnets, surface roughening, and melting~\cite{40-years-of-BKT}.
Theoretical understanding of the BKT mechanism is provided by RG theory
\cite{Kosterlitz}. Notable signatures of the BKT transition are the
universal jump of the superfluid density \cite{Nelson}, the
exponentially diverging correlation length, and the nonlinear
current-voltage characteristics of ultrathin superconducting 
films \cite{KTreview,Minnhagen}.

In this paper we focus on a less frequently studied quantity, the
magnetic permeability.  We study the scaling properties of the
permeability at the BKT transition by Kosterlitz renormalization group
theory and Monte Carlo simulations.  Andersson and Lidmar
\cite{Andreas} have pointed out that the vortex fugacity is
dangerously irrelevant for the free vortex density in the
low-temperature phase.  Here we find that this also applies to the
magnetic permeability.  We further show that the fugacity enters as a
multiplicative logarithmic correction to scaling that makes the
permeability vanish at the transition in the thermodynamic limit.
These results provide useful
additional information that complements the characterization by
superfluid transport, and should in principle be experimentally
measurable.  A multiplicative scaling correction with a similar origin
is known for the two-point order parameter correlation function \cite{Kosterlitz} and
the XY magnetization susceptibility \cite{Hasenbusch}.

The BKT transition has been frequently studied in numerical
simulations of XY models and Coulomb gas (CG) models.  In this paper
we test our predictions by large-scale Monte Carlo (MC) simulations of
two-dimensional Coulomb gas models and find good agreement with
theory.  To locate the transition in the absence of symmetry breaking
is somewhat complicated since the average of the order parameter
vanishes across the transition and alternative methods must be
invoked.  The most common quantity considered in simulations of the
BKT transition is the superfluid stiffness,
or, in the CG language, the dielectric constant.
To estimate the location
of the transition from numerical data for finite systems a finite-size
scaling method has to be used.  Corrections to scaling are significant
at the BKT transition and need to be included in a finite-size scaling
analysis.  The standard approach due to Weber and Minnhagen
\cite{Weber} is to include the lowest-order additive logarithmic
correction to the universal jump of the superfluid stiffness and make
a $\chi^2$ fit of numerical data close to the transition to estimate
the critical temperature from the best fit to the universal jump
criterion.  This approach has been used in a large number of
simulation studies of the BKT transition; see Ref.\ \cite{Hasenbusch}
and references therein.

We show that the magnetic permeability $\mu$ is also a convenient quantity
for locating the transition.
We use a simplified version of the Weber-Minnhagen method, by means of
an intersection analysis to extract the critical temperature from the
finite size correction, without any need to use optimization to find
the best fit.  We focus on finite-size scaling of the magnetic
permeability or, equivalently, the net vorticity fluctuation,
which in the CG corresponds to the compressibility, and which is 
rendered finite by slight modification of the model.  Our
analysis clearly demonstrates the presence of a multiplicative log
correction in the finite size scaling of our MC data and verifies the
predictions of the theory.  Furthermore, we find that the next order
correction to scaling is substantial and demonstrate how it can be
effectively included in the finite size scaling analysis.
Applying the same analysis to the superfluid stiffness gives consistent
results.

The paper is organized as follows.  First we discuss scaling of the
magnetic permeability at the transition.  Then we turn to the two
dimensional Coulomb gas model and the modifications needed to
calculate the magnetic permeability.  Then we describe our Monte Carlo
method and the finite size scaling approach including lowest and
next-order scaling corrections.  Finally the simulation results are
presented and discussed.

\section{Coulomb gas model and permeability of superconducting films}

We start from a London model for vortex fluctuations in two dimensional 
superconducting films.
The Hamiltonian is
\beq
H_{\theta}=\int d^2r \left[
\frac{J}{2} 
\left(\nabla \theta-\frac{2\pi \bA}{\Phi_0}\right)^2
+\frac{\bB^2}{2\mu_0} -B_zh \right]
\eeq
Here the coupling constant $J$ is the superfluid stiffness,
$\theta$ is the phase of the superconducting order parameter, 
$\bA$ is the vector potential, $\bB=\nabla\times\bA$ the 
magnetic flux density, $\Phi_0$ the flux quantum, 
$h$ is an applied perpendicular magnetic field,
and $\mu_0$ is the permeability of free space.
Vortices are topological defects in the phase field given by
\beq
\nabla\times\nabla\theta(\br)=2\pi \sum_n q_i \delta(\br-\br_n)
\eeq
By following standard steps the model can be reformulated and 
expressed directly in the vortex coordinates \cite{Lidmar}.  
The partition function then becomes 
\beq
Z_\theta = \int D\theta D\bA e^{-\beta H_\theta} \propto \sum_{N=0}^\infty \frac{z^N}{N!} \int \prod_{n=1}^N d^2 r_n \sum_{q_n=\pm 1} e^{- H/T} = Z_{CG}
\eeq
where the Hamiltonian for the vortices
\beq
H=\frac 12 \sum_{mn} q_mV(\br_m-\br_n)q_n - \Phi_0 h\sum_m q_m 
\eeq
takes the form of a CG model with interacting charges $q_n$ representing the vorticity.
Here $T = 2\pi J/\beta$ is the temperature in the CG model, $V(\br)$ is a screened Coulomb potential and 
the net vortex density equals the flux of the $B$-field in units of $\Phi_0$.
The fugacity $z=e^{-E_c/T}$ controls the vortex density, where $E_c$ is the vortex core energy.
For real 2d films the screening length corresponds to the Pearl length 
$\lambda=2\lambda_0^2/s$ where $s$ is the film thickness \cite{Minnhagen}. 
$\lambda$ can be of the order of centimeters
and can thus often be taken to be infinite, resulting in the usual 
CG model. For zero applied field $h$ and infinite screening length $\lambda$
the CG model undergoes a KT transition.

For finite screening length the BKT transition is in principle destroyed and replaced by a crossover.
In real superconducting ultrathin films the screening length is finite
but typically very large, comparable or exceeding the system size, meaning that the transition will still appear very sharp.
The finite screening length implies 
that fluctuations in the magnetic flux density are present, 
leading to a finite magnetic permeability given by
%
%
%
\beq
\mu = \frac{\partial \<B\>}{\partial h} = \frac{L^d}{T}(\<B^2\>-\<B\>^2) ,
\eeq 
where $d=2$ is the dimension.
In terms of the
flux of the $B$-field through a surface of area $L^2$ the total net vorticity is
$m=L^2B/\Phi_0$, and the permeability becomes
\beq 
\mu=\frac{L^{d-4} \Phi_0^2}{T}(\<m^2\>-\<m\>^2) ,
\label{permeability}
\eeq
In the CG language this corresponds to the compressibility
$\kappa = (1/L^d T)(\<m^2\>-\<m\>^2)$.
From now on we consider the zero field case with $\<B\>=\<m\>=0$.  
Below we reformulate the CG model on a lattice and study the permeability
with Monte Carlo simulation.

\section{Scaling results}

In 3d bulk type-II superconductors the magnetic permeability changes from zero 
in the Meissner phase to a finite value above the transition
and shows scaling behavior at the transition.
The scaling of the permeability from vortex fluctuations 
with system size $L$ in $d$ dimensions follows from a simple power counting argument.
The flux density scales as
$B=\nabla\times A\sim L^{-2}$~\cite{FFH}, 
the magnetic field as $h \sim L^{2-d}$ since $B h \sim L^{-d}$, and the permeability thus scales as 
$\mu \sim L^{d-4}$.
From Eq.\ (\ref{permeability}) this means that the 
net vortex fluctuation scales as
\beq
\<m^2\>\sim L^0
\label{mscal}
\eeq
Thus, in general the permeability is expected to vanish as a power law
at criticality and the net vorticity fluctuations to approach a constant
at the transition.  However,
as will be discussed next this turns out not to be the case at the
lower critical dimension $d=2$.

To construct corrected scaling relations that apply for $d=2$ a more
accurate treatment is needed of vortex interactions and vortex number
fluctuations.  The scaling properties are described by renormalization
group (RG) theory.  The RG flow for the Coulomb gas is most easily
expressed in terms of reduced temperature and fugacity variables
defined as $x=1-1/4T,y=\pi z/2T$.  The lowest order flow
equations are \cite{Kosterlitz}
\beq
\frac{dx}{dl}=2y^2 
\;,\;
\frac{dy}{dl}=2xy
\eeq
where $l=\ln b$ and $b$ is a rescaling factor.  The resulting RG flow
obeys
\beq
x^2-y^2=C^2
\eeq
where the constant $C$ is determined by the initial conditions.  Below
$T_c$ we have $C^2>0$ and above $T_c$ we have $C^2<0$.  The BKT
transition occurs for $T=T_c$ and then the flow obeys $x+y=0$.
Explicit solutions are given by
\beq
y=\left\{
\begin{array}{ll}
	\dfrac{2C(b/b_0)^{-2C}}{1-(b/b_0)^{-4C}} & \mbox{ for } T<T_c \\
	\dfrac{1}{2\ln b/b_0} & \mbox{ for } T=T_c 
	\\
	- \dfrac{|C|}{\sin [2|C|\ln(b/b_0)]} & \mbox{ for } T>T_c 
\end{array}
\right.
\eeq
where $b_0$ is determined by the initial condition for the RG flow.  
At and below the transition, the RG flow ends on the critical line $x\leq 0,\, y=0$.

According to the naive scaling Eq.~(\ref{mscal}), 
the mean square vortex density should approach a constant for $b\to\infty$.
This is not correct at the BKT transition where instead
the mean square vorticity is expected to be proportional to the
renormalized fugacity that scales to zero as $b\to\infty$.
Consider the renormalization of the magnetic permeability,
\begin{equation}\label{eq:renorm-mu}
\mu(x,y,L, \ldots) = b^{d-4} \mu(x(b), y(b), L/b, \ldots)
\end{equation}
The naive scaling would hold only if the right hand side tends to a
constant as $b\to\infty$, but this is not the case when $y\to 0$,
i.e., at and below $T_c$.
This is seen by explicit calculation of the partition function for $m=0,\pm 1$
which yields for small $z$
\beq
\<m^2\> \approx 2z \sim y
\eeq
and thus goes to zero at the fixed line.
For the permeability this gives $\mu \sim y/L^2$.  This modifies the
naive scaling result $\<m^2\>\sim constant$ and demonstrates that the
fugacity gives a multiplicative correction to scaling.  The 
multiplicative scaling correction makes the net vorticity 
asymptotically approach zero at the BKT transition.
Stopping the RG flow at $b \approx L$ gives the finite size scaling formulas
\begin{equation}\label{eq:fss}
\mu(L) \sim L^{-2} y(L) \sim
\left\{
\begin{array}{ll}
	L^{-2 + 2x_R} & \mbox{ for } L \gtrsim \xi_- \\
	\dfrac{L^{-2}}{\ln L/b_0} & \mbox{ for } L \lesssim \xi_-  
\end{array}
\right. 
\end{equation}
where $\xi_- \approx \exp(1/2C) \approx \exp(1/2c\sqrt{T_c-T})$
is the correlation length below $T_c$,
and $x_R = x(b \to \infty) = 1-1/4 T_R$.
Right at $T_c$ this gives a multiplicative logarithmic correction,
while below $T_c$ the naive power-law scaling is changed into one with a 
temperature dependent exponent.
 
The rest of the paper will study these relations by finite size
scaling of data from Monte Carlo simulation.  In the analysis of MC
data the scale factor $b$ will be taken to be the finite system size
$L$, and the initial condition $b_0$ is an UV cutoff that corresponds
to the vortex core radius $L_0$.

\section{MC simulation of the lattice Coulomb gas}

To test the modified scaling relations described above we performed
large-scale Monte Carlo simulations of a Coulomb gas model for vortex
fluctuations.  The lattice two-dimensional Coulomb gas (CG) model is
defined by~\cite{Lee-Teitel1990,Lee-Teitel1992}
\beq
H=\frac{1}{2} \sum_{i,j} q_iV_{ij}q_j-\mu_v N
\label{CG}
\eeq
where $q_i=0,\pm 1,...$ is the Coulomb gas charge, or equivalently
vorticity, on lattice site $i$ of a square lattice with $L\times L$
sites with periodic boundary conditions.  Here we only consider the
case of no net applied magnetic field.  The lattice Coulomb
interaction is given by
\beq
V_{ij}=\frac{2\pi}{L^2} \sum_{\bk}
\frac{e^{i\bk\cdot\br_{ij}}}{4\sin^2 (k_x/2)+4\sin^2 (k_y/2)}
\label{interaction}
\eeq
where $k_\mu=2\pi n_\mu/L$, and we set the lattice spacing to $a=1$.
$\mu_v$ is the vortex chemical potential, $E_c=-\mu_v$ is the vortex
core energy, and $z=e^{\beta \mu_v}$ is the fugacity.  The particle
number or total vorticity is $N=\sum_i |q_i|=N_++N_-$, and the net
vorticity is $m= \sum_i q_i=N_+-N_-$.  The partition function is
\beq
Z=\sum_{\{q_i\}} e^{-\beta H}
\label{Z}
\eeq
where $\beta=1/T$.  We studied a few different values of the vortex
chemical potential $\mu_v$.  All results shown below are for
$\mu_v=0$.  The other values of $\mu_v$ that we investigated gave
similar results.

The CG model described above has $V(k=0)=\infty$, which means that
fluctuations in the net charge cost infinite energy and are excluded.
Thus the net vorticity is restricted to $m=0$ which makes $\<m^2\>=0$.
Therefore, to simulate the effect of fluctuations of the net vorticity
in order to enable calculation of the permeability given by Eq.\
(\ref{permeability}) from $\<m^2\>$, the CG model has to be modified.
This can be done in different ways.  Including fluctuations
in the perpendicular $B$-field leads to a CG model in Fourier
space given by
\beq
H=\frac{1}{2L^2}\sum_{\bk} |q(\bk)-B(\bk)|^2 V(\bk) + \lambda^2 |B(\bk)|^2
\eeq
where $\lambda$ is the screening length.
Here $B(\br)$ correspond to the magnetic flux through a plaquette in units of $\Phi_0$.
Since $V(k=0)=\infty$,
uniform fluctuations in $B(\bk=0)$ are accompanied by fluctuations in
the net vorticity such that $q(k=0)=B(k=0)$.  Integrating out the
fluctuations in $B(\bk)$ leads to a screened Coulomb interaction given
by
\beq
V_{ij}=\frac{2\pi}{L^2} \sum_{\bk}
\frac{e^{i\bk\cdot\br_{ij}}}{4\sin^2 (k_x/2)+4\sin^2 (k_y/2)+\lambda^{-2}}
\label{scrinteraction}
\eeq
For finite $\lambda$ the self energy is finite and therefore charged
configurations appear so that $\<m^2\>$ can be studied.  We considered
two different models of screening.  The first model includes a finite
screening length for the $\bk=0$ term only, that corresponds to only
including fluctuations in the uniform part of $B$.
In this sense this corresponds to a minimal modification of the unscreened model.
In the second
model $\lambda$ is finite for all $\bk$ which corresponds to including
fluctuations in all $B(\bk)$.

The value of the screening length $\lambda$ needs to be selected in a
special way to guarantee that the CG has a BKT transition.  The
problem is that the screened models described above have fluctuations
in the net vorticity present at all temperatures.  This means that the
low-temperature superconducting phase where vortices are present only
in neutral dipole pairs is destroyed, and the system is always in the
high-temperature phase.
In the RG sense screening is a relevant perturbation, with the screening 
length scaling as $\lambda \sim b$.
Thus, for finite $\lambda$ the BKT transition will be replaced by crossover.
To circumvent this problem and
retain a BKT transition
we define the thermodynamic limit by taking the screening length
proportional to the system size, i.e.,
$\lambda=cL$.  Then, for $L\to\infty$ the screening length diverges
and the system has a BKT transition in the thermodynamic limit, and
furthermore has {\it the same} $T_c$ as the unscreened CG model
since the scaling combination $\lambda^{-1}L = c^{-1}$ is held
constant in all scaling functions.  This
scheme at the same time includes magnetic field fluctuations and
retains a BKT transition that can be studied by finite size scaling.
We simulated both the models described above of magnetic field
fluctuations for different choices of the constant $c$, and obtained
similar results.  Results below are shown for the model with only uniform
$B$-field fluctuations for the choice $\lambda=L/2$.

The main focus of this paper is to study how the BKT transition is
seen in the finite size scaling properties of MC data for various
quantities.  For the neutral case with $m=0$ (and $\lambda=\infty$), the BKT transition can
conveniently be located from simulation data using the
Nelson-Kosterlitz universal jump \cite{Nelson} of the superfluid
density $\rho_s$ at the transition.  In the CG language this
corresponds to a universal jump in the dielectric response function
$1/\epsilon=2\pi\rho_s/\rho_0$ that will be considered here.  At the BKT
transition temperature $T_c$ the dielectric response function $\epsilon^{-1}$ jumps
from the finite value
\beq
\frac{1}{\epsilon} = 4T_c
\label{univjump}
\eeq
to zero at the transition.  To calculate the dielectric function in a
MC simulation of the CG it is useful to add a polarization term to the
energy \cite{Olsson1995,Lidmar1997}
\beq
H=H_0+\frac{\pi}{L^2}\bP^2
\eeq
where $\bP=(P_x,P_y)=\sum_i q_i \br_i$ is the polarization.  The
dielectric function is then obtained from the polarization fluctuation
by \cite{Olsson1995,Lidmar1997}
\beq
\frac{1}{\epsilon}=1-\frac{\pi}{L^2T} (\<\bP^2\>-\<\bP\>^2)
\label{epsilon}
\eeq

The Monte Carlo (MC) simulation generates a Markov chain of vortex
configurations on the lattice by repeating the following trial moves
of inserting charges.  The initial configuration is taken to be an
empty system with no charges and $N=0$.  The code uses two kinds of MC
trial moves that are accepted with the Metropolis acceptance
probability $P=\mbox{min} (1,e^{-\beta\Delta H})$.

The first kind of MC trial move attempts to insert a neutral dipole
pair of charges with $q=+1,-1$ on a randomly chosen nearest neighbor
pair of lattice sites.  This move automatically takes care of both
creation, destruction, and movement of dipole pairs.  Adding a neutral
pair will not change the net charge of the system.  This is the only
type of MC move used in the simulation of the CG in the neutral case
where $V(k=0)=\infty$ and $m=0$.

The second kind of MC move attempts to add a single charge with a
random sign, $q=\pm 1$, generated with equal probability to a randomly
chosen lattice site.  The move changes the net charge $m$ of the
system and is only used in the case where $V(k=0)\ne 0$.  Each MC move
is randomly chosen with equal probability to be of the first or second
kind.

We refer to one sweep through the system as $L\times L$ update
attempts to insert dipoles or single charges.  Other types of MC moves
like moving or removing particles or pairs can also be used and
potentially improves convergence properties but we settled with the
moves described above since they gave satisfactory convergence of the
simulation.  We found that $10^3$ initial sweeps to establish
equilibrium was sufficient.  In equilibrium measurements of the
observables were done after each sweep.  A total of $10^4$ terms were
collected to form averages.  The runs were repeated about $10^2$ times
until sufficiently small statistical errors had been obtained.  Error
bars were estimated from the standard deviation of the results from
different runs.  Single histogram reweighting was used to obtain data
at a range of nearby temperatures from simulations done at $T=0.2115$
\cite{numrec}.

\section{Results}

As a first step we discuss how the BKT transition temperature 
can be estimated from MC data using the universal jump criterion.  
In this calculation we use the CG model without any fluctuations in the 
net vorticity so that $m=0$ throughout the simulation.
Figure 1 shows MC data for $1/\epsilon$ obtained by evaluating
Eq.\ (\ref{epsilon}) for a sequence of different system sizes $L=2^n=4,8,16,...,256$
and temperatures around $T_c$ obtained from single-histogram reweighting.
According to the universal jump prediction, Eq.\ (\ref{univjump}) the
data points should jump from a system size independent value
$1/\epsilon=4T_c$ to 0 at the BKT transition.  But due to the presence
of big finite size effects, no clear indication of a jump is seen in
the data for finite system sizes.  Instead, as the system size
increases we expect the data curves to slowly approach the
characteristic square root cusp at $T_c$ and undergo the universal
jump \cite{KTreview}.  As seen in the figure the approach to the
asymptotic behavior is very slow which complicates the estimation of
$T_c$.

\begin{figure}
	\includegraphics[width=0.8\columnwidth]{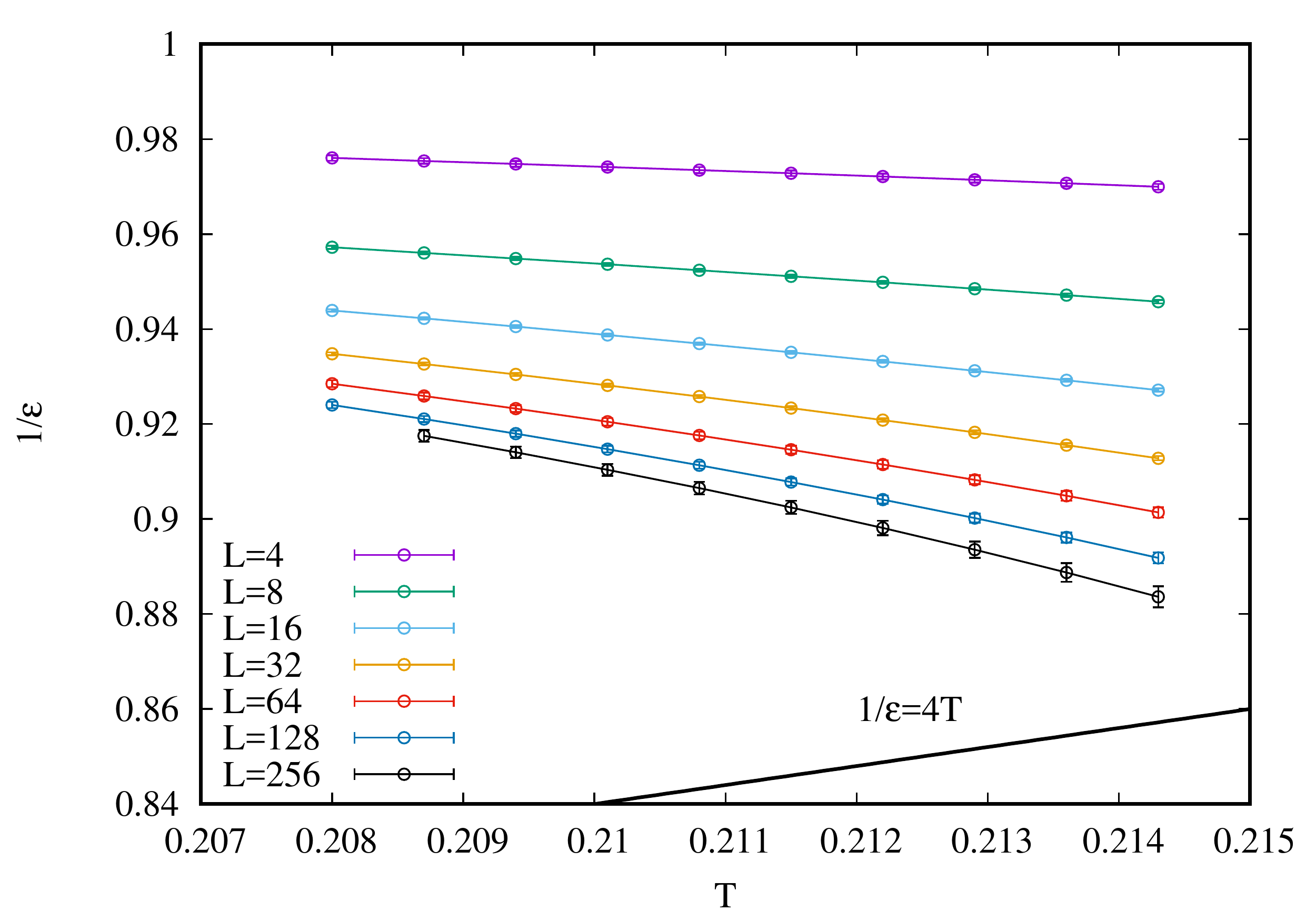}
	\caption{MC data for the dielectric function vs temperature of
          the CG model without screening ($\lambda=\infty$) for
          different system sizes $L$ showing substantial finite-size
          effects. The solid line represents the universal jump
          $1/\epsilon=4T$. The transition temperature is where a data
          curve for $L\to \infty$ would intersect the straight black
          line.}
	\label{fig1}
\end{figure}

\begin{figure}
	\includegraphics[width=\columnwidth]{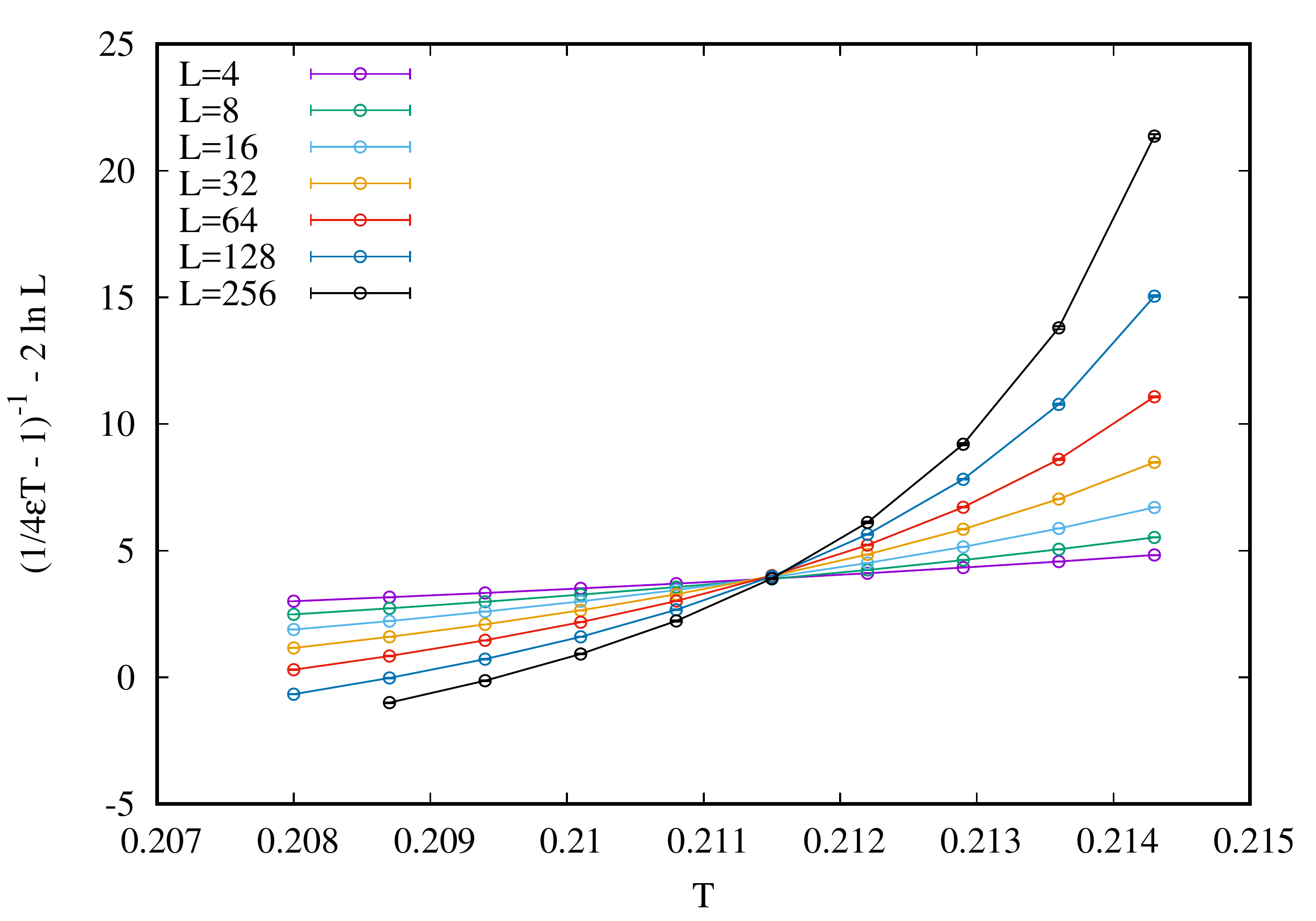}
	\caption{Intersection plot of MC data for the dielectric
          response function including the logarithmic correction to
          scaling according to Eq.\ (\ref{intersection}).  The BKT
          transition is located at the intersection point
          $T_c=0.2115\pm 0.0001$.}
	\label{fig2}
\end{figure}

The problem with the slow approach to the universal jump is overcome
by the Weber-Minnhagen finite size scaling form of the approach to the
universal jump given by \cite{Weber} 
\beq
\frac{1}{4\epsilon T_c}=1+\frac{1}{2\ln L+C}
\label{weberscaling}
\eeq
where $C=-2\ln b_0$ is an unknown constant.  This form follows from
Kosterlitz RG equations and gives the leading additive logarithmic
finite-size correction to the universal jump value.  A common approach
to estimating the transition temperature $T_c$ is to minimize the RMS
deviation between numerical data and the finite size form over
variations in both $T_c$ and $C$.  Here we propose a simpler method.
If the universal jump value is assumed to be correct, solving
Eq.~(\ref{weberscaling}) for the constant $C$ gives
\beq
C=\frac{1}{1/4\epsilon T_c-1}-2\ln L .
\label{intersection}
\eeq
A plot of MC data curves for $\frac{1}{1/4\epsilon T-1}-2\ln L$ vs $T$
for different system sizes $L$ is thus
expected to produce a system-size independent intersection point at
$(T_c,C)$.  This procedure involves no parameter fitting and
straightforwardly produces an accurate estimate.  Figure \ref{fig2}
displays an intersection plot according to Eq.\ (\ref{intersection}).
The intersection point is at $T_c = 0.2115\pm 0.0001$ and $C=4.0$,
giving $b_0 \approx e^{-C/2} \approx 0.135$.
Since a sharp intersection point is obtained, the universal jump
assumption is confirmed.  For the smallest lattice sizes a small
deviation from the intersection point is visible which indicates the
presence of higher order corrections to scaling.  The assumption made
here of the size of the universal jump is actually not necessary since
a similar intersection method described below to test this result is
straightforward.

We now turn to calculation of the vorticity fluctuation.  The main
result of this paper is that a useful finite size scaling approach to
the BKT transition is offered by the magnetic permeability which is
proportional to the fluctuation in the net vorticity $\<m^2\>$.
Calculating $\<m^2\>$ requires a modification of the CG model since
the unscreened Coulomb potential does not permit charged
configurations with nonzero net vorticity.  In order to have nonzero
values of $\<m^2\>$ we simulate a modified CG model with interaction
given by Eq.\ (\ref{scrinteraction}) that includes
magnetic field fluctuations.
This model is expected to have the same
thermodynamic critical temperature as estimated above,
since the thermodynamic limit is approached by taking $\lambda \propto L$.

\begin{figure}
	\includegraphics[width=\columnwidth]{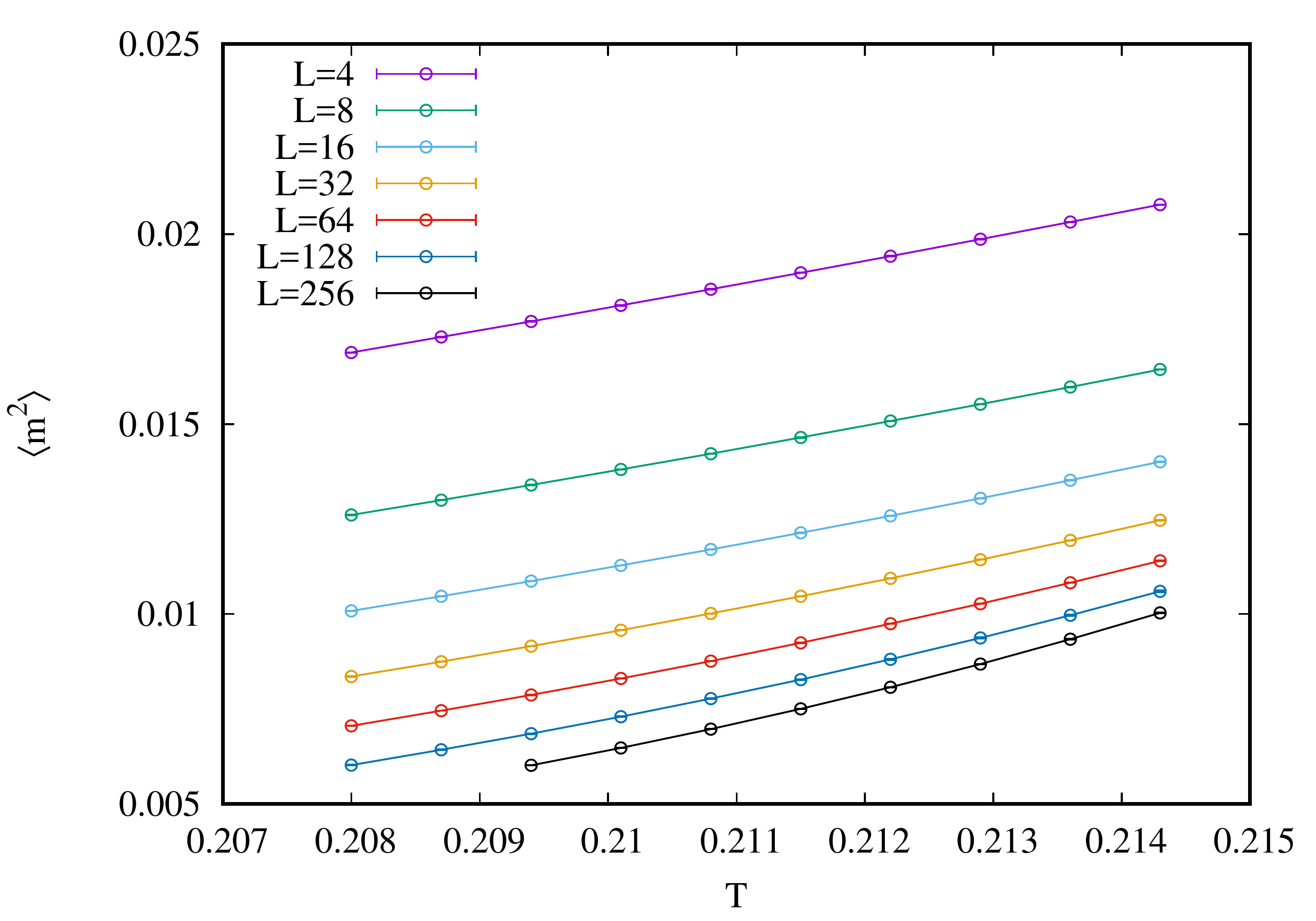}
	\caption{MC data for the net vorticity fluctuation vs
          temperature close to the BKT transition at $T_c\approx 0.2115$.}
	\label{fig3}
\end{figure}

Figure \ref{fig3} shows MC data for $\<m^2\>$ vs $T$ for different
system sizes $L$.  According to the naive scaling result
$\<m^2\> \sim L^0$ at $T=T_c$ the data curves should intersect at a
single point in the plot.  This is not the case demonstrating the
significance of including scaling corrections, similar to the case
above with the dielectric response function $1/\epsilon$.  As
discussed above the fugacity is expected to be a dangerous irrelevant
variable producing a multiplicative scaling correction for the
vorticity fluctuation and we next turn to identifying this correction
in simulation data.

\begin{figure}
	\includegraphics[width=\columnwidth]{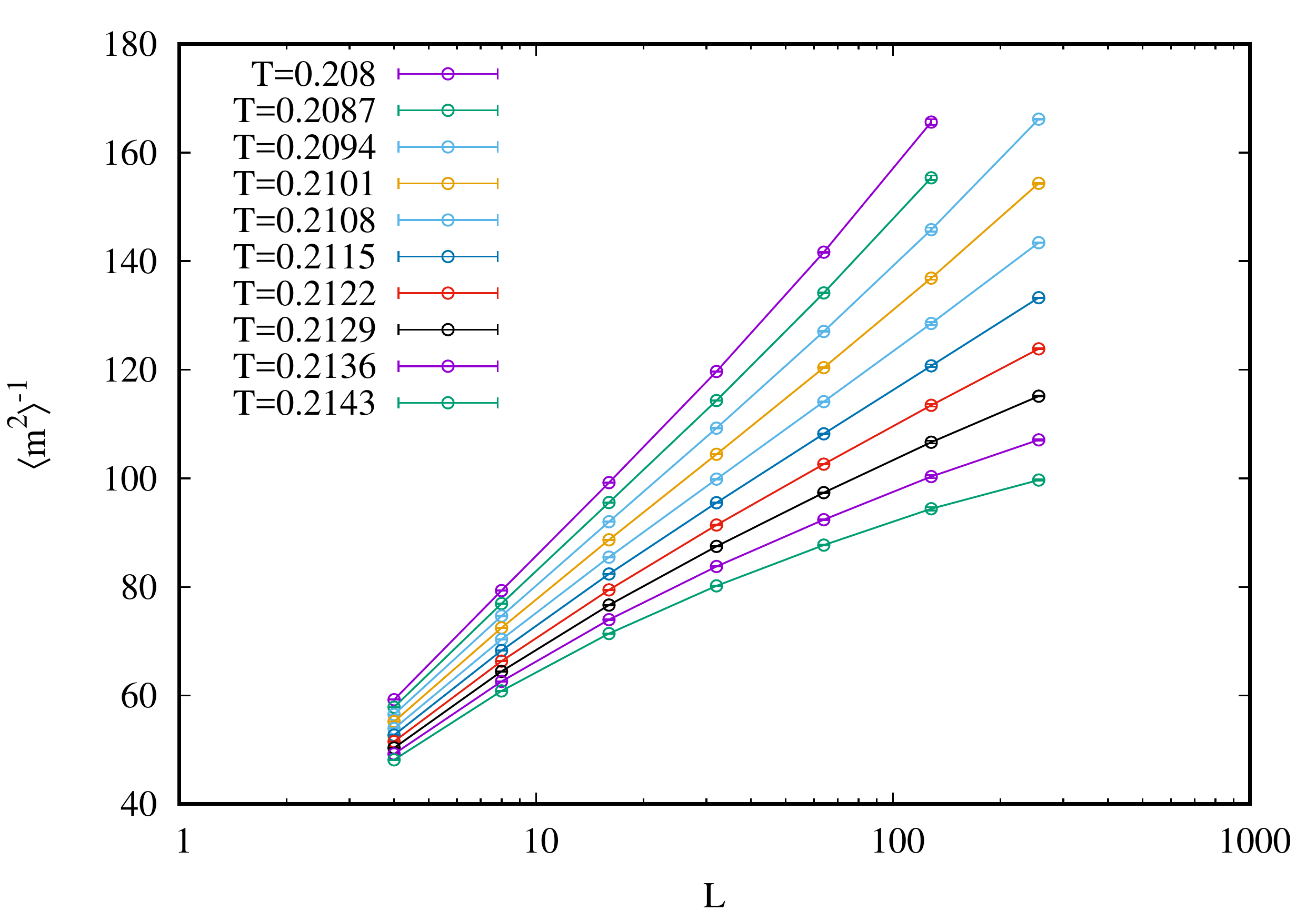}
	\caption{Inverse vorticity fluctuation vs system size $L$ for
          different temperatures.  According to Eq.\~(\ref{m2inverse})
          the BKT transition is where the data curves approach
          $\sim \ln L$ forming a straight line for large $L$.  Away
          from the transition the curves splay out away from the
          $\ln L$ form. }
	\label{fig4}
\end{figure}

According to Eq.\ (\ref{eq:fss}) the inverse of the vorticity
fluctuation should scale as
\beq
1/\<m^2\>=A\ln L + B
\label{m2inverse}
\eeq
at $T=T_c$, where $A,B$ are constants.  Figure \ref{fig4} shows MC
data for $1/\<m^2\>$ vs $\ln L$.  While an indication of a linear
dependence on $\ln L$ is obtained at $T\approx 0.2115$ for large $L$,
it is not possible in this figure to accurately estimate $T_c$.  One
possibility is to proceed to fit the MC data to a straight line and
estimate $T_c$ by minimizing the fit error at large $L$.  Instead we
again prefer an intersection method that eliminates the need for
fitting.

\begin{figure}
	\includegraphics[width=\columnwidth]{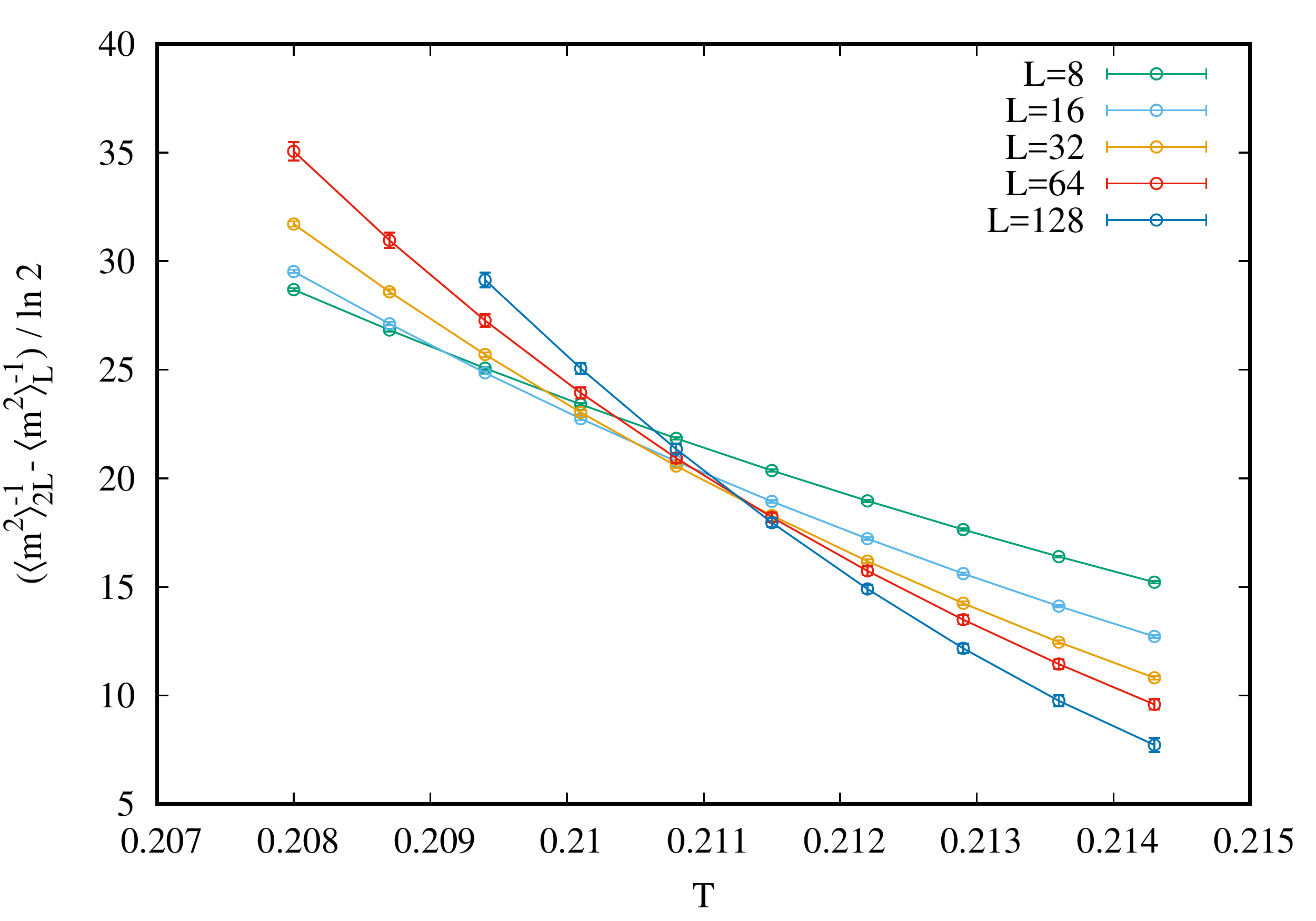}
	\caption{Intersection plot of MC data for $\<m^{2}\>^{-1}$ vs
          temperature.  The drift away from a common intersection
          indicates the presence of substantial finite size effects
          corresponding to higher order corrections to scaling.}
	\label{fig5}
\end{figure}

An intersection method for data for $\<m^2\>$ works in the following
way.  The constant $B$ in Eq.\ (\ref{m2inverse}) is eliminated by a
subtraction using pairs of MC data points with system sizes $(2L,L)$
to form
\beq
(\<m^2\>^{-1}_{2L}-\<m^2\>^{-1}_L)/\ln 2=A
\eeq
where subscript $L$ denotes data for system size $L$.  Hence this
quantity is expected to be independent of system size at the
transition.  Figure \ref{fig5} shows the corresponding MC data curves.
An intersection point for large system sizes is found at
$T_c \approx 0.2115$ which agrees with the value obtained for the
dielectric function in Fig.\ \ref{fig2}.  However, the intersection
quality is not nearly as good and finite size effects are substantial
despite including the correction to scaling.  In a sense it is not
unexpected that the approach to the transition is slower in this case
than in the neutral case since we introduced another length scale 
$\lambda \propto L$ in
the problem.  In addition the intersection point gives the estimate
$b_0 \approx e^{-B/A} \approx 0.151$, which is similar to the value $0.135$ found from
dielectric constant above.
Also note that a similar subtraction method enables a direct test
of the size of the universal jump in the dielectric function.

\begin{figure}
	\includegraphics[width=\columnwidth]{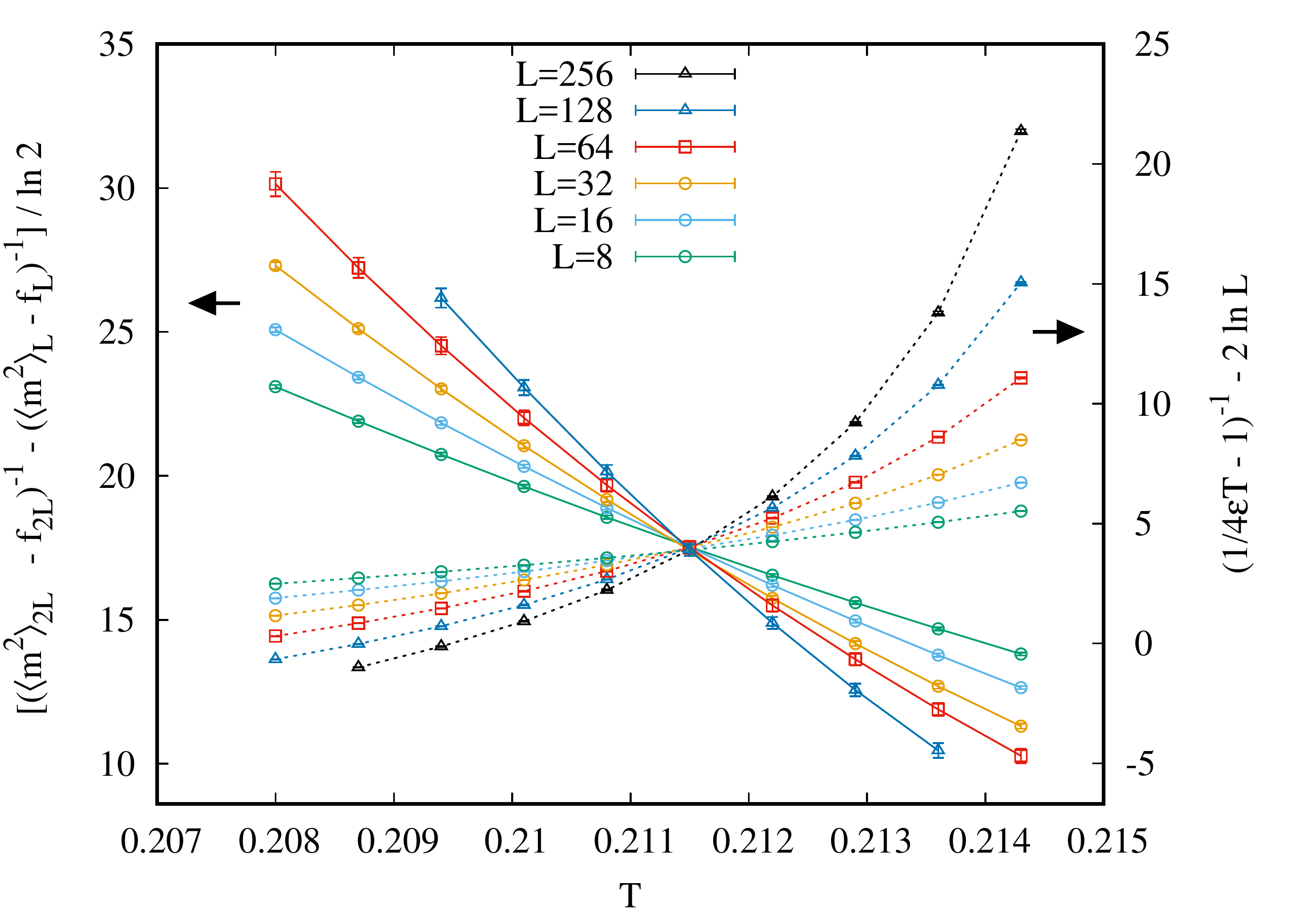}
	\caption{Solid curves: Intersection plot of MC data for a
          quantity involving subtraction of data for $\<m^2\>$ for
          pairs of system sizes $(2L,L)$, and scaling corrections up
          to second order given by $f=C\ln\ln L/\ln^2 L$. The data
          curves intersect at a single point for $C=-0.02$.  Dotted
          curves: For comparison the intersection curves for the
          dielectric function from Fig.\ \ref{fig2} are included.  A
          {\it common} sharp intersection point is produced for both
          quantities at the transition temperature $T_c=0.2115$.}
	\label{fig6}
\end{figure}

Since the intersection points in Fig.\ \ref{fig5} drift with system
size, it is motivated to attempt to include the next order correction
in the finite size scaling analysis.  The form of the next order
correction for large length scales is known and for the dielectric constant
becomes \cite{Amit,Hasenbusch,Sandvik}
$1/4\epsilon T_c=1+1/2\ln L/L_0 + {\rm const} \cdot \ln \ln L/\ln^2
L$.  Adopting this form of the correction to the charge fluctuation at
$T_c$ gives
\beq
\<m^2\>=\frac{1}{A\ln L + B} + C\frac{\ln \ln L}{\ln^2 L}
\label{first}
\eeq
where $A,B,C$ are unknown constants.  Compared to the case of fitting
the dielectric function where the known universal jump could be used,
the charge fluctuation involves one more unknown constant, which
complicates fitting to numerical data.  A possible approach is to look
for the parameters that give the best $\chi^2$-fit to the data.
Instead we construct a simple intersection plot that contains the same
information.  Equation (\ref{first}) gives
\beq
(\<m^2\>-f_L)^{-1}=A\ln L + B
\label{second}
\eeq
where $f_L=C\ln\ln L/\ln^2 L$.  Subtraction of this quantity for pairs
of system sizes $(2L,L)$ eliminates $B$ and gives
\beq
A=\left[(\<m^2\>_{2L}-f_{2L})^{-1}-(\<m^2\>_L-f_L)^{-1}\right]/\ln 2
\label{third}
\eeq
Hence a plot of this quantity is expected to produce a system size
intersection at $T_c$ when the optimal value of $C$ is used.  A
straightforward optimization locates the best intersection point with
minimal scatter between curves with different system size $L$ and
produces the estimates 
$T_c=0.2115\pm 0.0001, A=17.47\pm 0.02, B=19.17\pm 0.08, C=-0.02$.
The resulting intersection plot is shown in Fig.\ \ref{fig6}.  The
solid lines are data curves for the charge fluctuation, and the dotted
curves are data for dielectric function from Fig.\ \ref{fig2}.
Evidently both quantities produce sharp intersections at precisely the
same temperature.

\begin{figure}
	\includegraphics[width=\columnwidth]{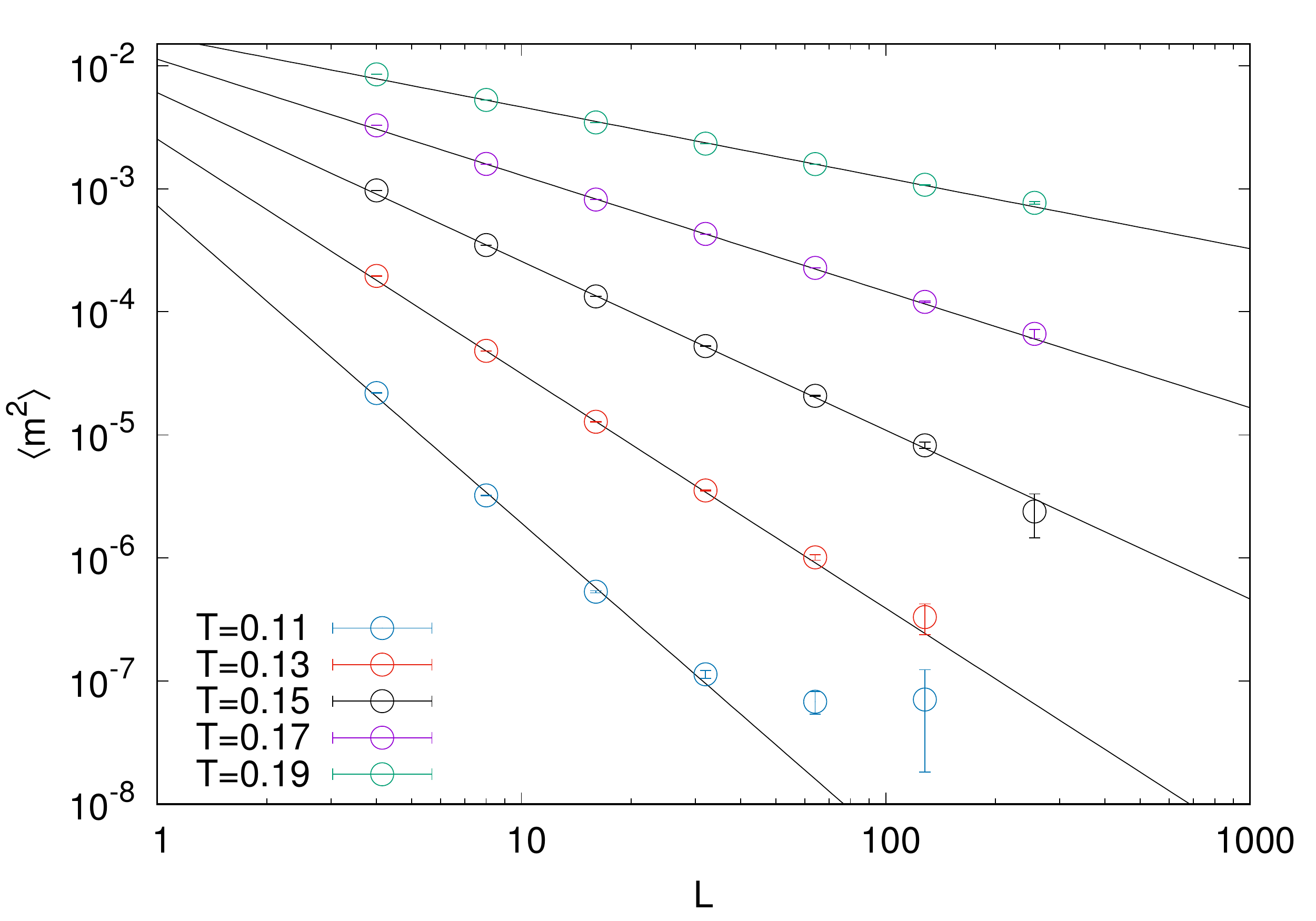}
	\caption{MC data for $\<m^2\>$ vs $L$ for different temperatures below $T_c$.
	The solid lines are power law fits.}
	\label{fig7}
\end{figure}

\begin{figure}
	\includegraphics[width=\columnwidth]{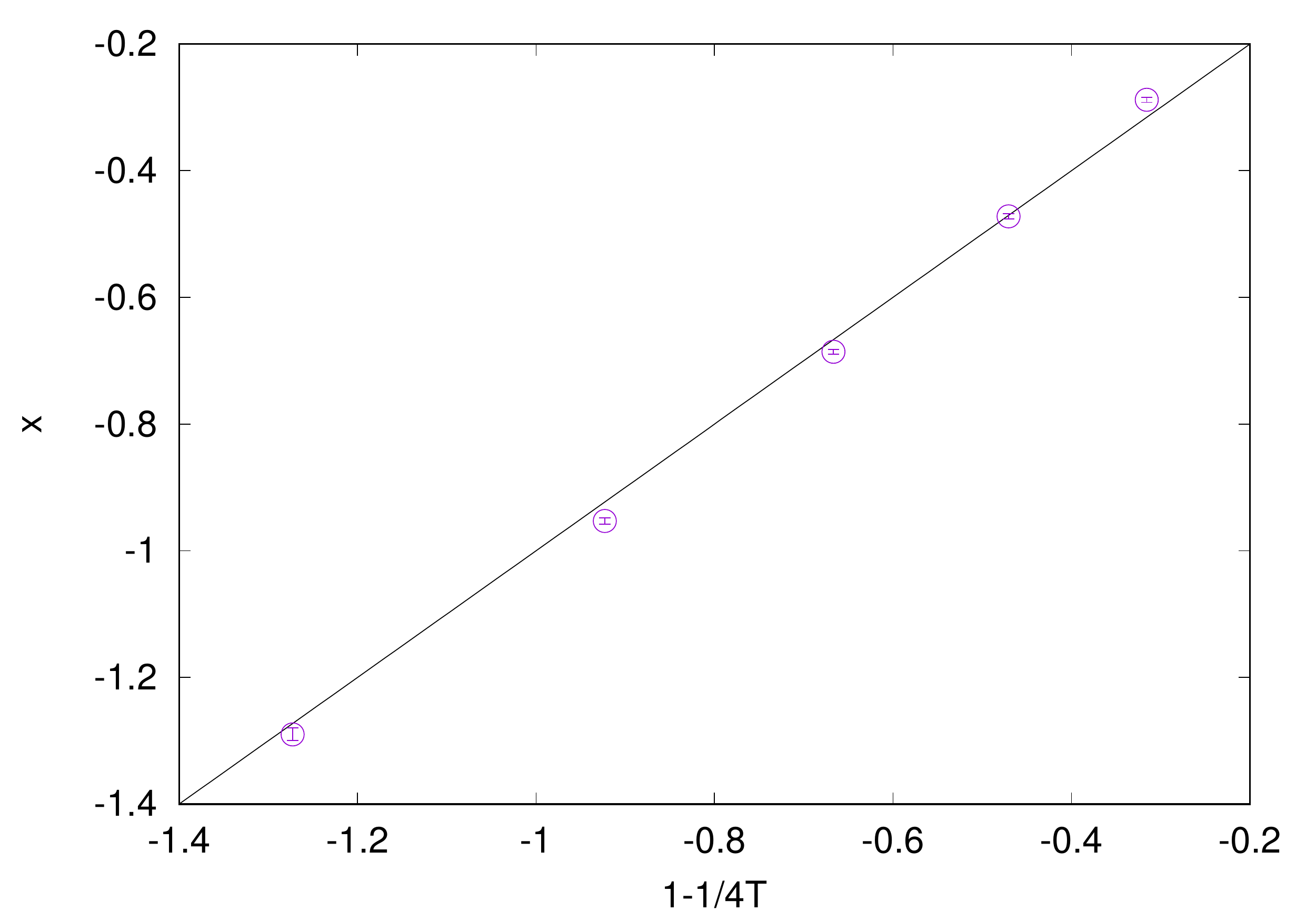}
	\caption{Power law exponent $x$ in $\<m^2\>\sim L^{2x}$ obtained 
		from the fits in Fig.\ \ref{fig7}.  The points are close to 
		the straight line which is expected from the predicted power 
		law dependence with $x \approx 1-1/4T$.}
	\label{fig8}
\end{figure}

Finally we investigate the scaling properties of $\<m^2\>$ below $T_c$.  
According to Eqs.\ (\ref{permeability}),(\ref{eq:fss}) the size dependence 
here becomes a power law with a temperature dependent exponent given by
$\<m^2\> \sim L^{2x_R(T)}$. Figure \ref{fig7} shows MC data points for $\<m^2\>$ 
plotted vs $L$ for temperatures $T<T_c$.  
The solid lines are power law fits to the MC data points.  Figure \ref{fig8}
shows the fitted power law exponent $x$ plotted vs $1-1/4T$.
The theoretical prediction discussed above is $x_R(T)=1-1/4T_R$ 
where $T_R$ is the renormalized temperature.
Except very near $T_c$ we expect $T_R \approx T$, $T_R \geq T$.
Since the data in the figure falls nearly on a straight line the 
agreement with the predicted power law is quite plausible.

\section{Discussion}

BKT physics is usually studied theoretically in neutral vortex systems where vortex fluctuations enter only as neutral dipole pairs that dissociate at the BKT transition.
Here we consider BKT physics in non-neutral vortex ensembles and show that fluctuations in the net vortex number is a most useful quantity for studying the BKT transition.  In finite systems such non-neutral fluctuations cost finite energy and are therefore present and can in principle be measured which makes our predictions relevant for experiments.  In superconductors this corresponds to measurements of the magnetic permeability.  In the usual CG model on a finite system with periodic boundary conditions that we study by MC simulations in this paper, the vortex system is incompressible due to the infinite Coulomb self energy, and the net vorticity is always zero.  To include net vortex fluctuations and at the same time not eliminate the BKT transision we introduce the useful trick of a system size dependent screening length $\lambda\sim L$ that makes the single-vortex energy finite.
This introduces net vortex fluctuations at all temperatures and permits calculation of the compressibility.  
At the same time the screening length diverges with system size so in the thermodynamic limit the BKT transition is intact and located at the same critical parameters as the unscreened CG model.  
This technique gives a useful route for studying phase transitions by 
circumventing a sum rule such as charge neutrality in the CG case.

We considered the scaling properties of the non-neutral vortex fluctuation and 
obtained useful finite size scaling results for the BKT transition.
The usual finite size scaling analysis of the universal jump in the superfluid density including the Weber-Minnhagen additive log correction to scaling
can conveniently be implemented in an intersection plot of MC data for different system sizes.  The intersection point directly estimates the BKT transition temperature.  For the analysis of charge fluctuations we show that the naive power counting scaling results are violated, and we construct the proper scaling relations from Kosterlitz RG theory. We obtained a multiplicative scaling correction to the power-counting scaling form.
Furthermore, the next-order scaling correction is needed to get an accurate estimate of the critical point and we implemented this calculation in an intersection analysis. We found excellent agreement between the different methods for locating the transition.
Finally we showed that Kosterlitz RG theory leads to a power-law dependence 
of the charge fluctuation with system size which is to a good approximation consistent with our MC data.

In summary we consider non-neutral vortex fluctuations as a useful quantity for studying the BKT transition and show that it has a novel multiplicative log correction to scaling at the transition.  It would be interesting to look for such properties in experiments by for example magnetic permeability measurements in effectively two-dimensional superconductors
and in cold atom systems.

\section{Acknowledgements}

We thank Hans Weber, Nikolay Prokof'ev and Boris Svistunov for useful discussions.
This work was supported by the Swedish Research Council VR grant
621-2012-3984.  Computations were performed on resources provided by
the Swedish National Infrastructure for Computing (SNIC) at HPC2N.

\bibliographystyle{apsrev4-1}
\bibliography{permeability.bib}
\end{document}